\documentclass[reprint,amsmath,amssymb,aps,prb,showkeys,]{revtex4-2}
\pdfoutput=1
\usepackage{graphicx}% Include figure files
\usepackage{dcolumn}% Align table columns on decimal point
\usepackage{bm}% bold math
\usepackage{subfigure}
\DeclareMathOperator{\sech}{sech}

\begin{document}

\preprint{APS/123-QED}

\title{Effect of Oxygen on Hydrogen Diffusivity in $\alpha$-Zirconium}

\author{Manura Liyanage}
\email{manuraliyanage@cmail.carleton.ca}
\affiliation{Department of Civil and Environmental Engineering,\\ Carleton University,\\ Ottawa,
ON K1S 5B6, Canada}
\author{Ronald Miller}
\affiliation{Department of Mechanical and Aerospace Engineering,\\ Carleton University,\\ Ottawa,
ON K1S 5B6, Canada}%
\author{R. K. N. D.  Rajapakse}
\affiliation{School of Engineering Science,\\ Simon Fraser University, Burnaby, BC V5A 1S6, Canada}%

\date{\today}% It is always \today, today,
             %  but any date may be explicitly specified

\begin{abstract}
Zirconium and its alloys are extensively used as cladding material in nuclear reactors. They are vulnerable to hydrogen degradation under the harsh service conditions of the reactors, which necessitates continuous monitoring for the hydride concentration.  The presence of hydride denuded zones in the latter stages of the pressure tube's life hinders the monitoring process, which is carried out by scrape samples taken from the surface of pressure tubes. We investigated the effect of oxygen on diffusivity of hydrogen in $\alpha$ Zr, to check the hypothesis that oxygen slows the diffusion of hydrogen and thereby encourages the occurrence of hydride denuded zones. From the study we found that oxygen indeed decreases the diffusivity of hydrogen in $\alpha$ Zr for moderate O concentrations, supporting the hypothesis. We investigated the diffusion processes of individual H atoms, which showed that the reduction in diffusivity is caused by a decrease in the hopping rates and the formation of hydrogen traps by the combination of several interstitial sites.
\end{abstract}

%Use showkeys class option if keyword
                              %display desired
\maketitle

%\tableofcontents

\section{\label{sec:introduction}Introduction}
Zirconium and it's alloys are used to make pressure tubes, which contains the nuclear fuel in the reactors, due to the low thermal neutron absorption, good mechanical properties, and good corrosion resistance of Zr alloys \cite{Krishnan1981,Banerjee2001,Feron2012}. In the harsh service conditions of nuclear reactors, these alloys react with water to form zirconium oxide, while releasing hydrogen. Some of this hydrogen enters the metal which adversely changes  its properties like the fracture toughness. Delayed hydride cracking can also occur if sharp flaws, cuts, and nicks suffered during operation coincide with regions under tensile stresses \cite{McRae2010,DeLasHeras2018,Murty2019}. Hence, the hydride content in these pressure tubes are continuously monitored through scrape samples taken from their surfaces.  Optical micrographs suggest that the size of hydrides close to the metal-oxide interface can be smaller than in the bulk which hinders the monitoring process \cite{Hardie1965,Cha2005}. While the reasons for these hydride ``depleted" zones are not exactly known, it is hypothesized that oxygen slows down the diffusion of H causing it to form smaller intragranular hydrides instead of the usual intergranular hydrides found as needle or plate shaped precipitates \cite{Hardie1965}. After an extensive literature review we didn't find any studies which has investigated this hypothesis through experimental or computational means.

We conducted a multi-scale study using first-principles calculations and Kinetic Monte Carlo (KMC) simulations to show the validity of this hypothesis by ascertaining the diffusivity of H in hexagonal close packed (hcp or $\alpha$ phase) Zr at zero to moderate O concentrations. We also looked at the reasons for the variation in H diffusivity due to the presence of O by looking at the diffusion mechanics and diffusion paths taken by H atoms.

\section{\label{sec:comp_details}Computational Details}
First principles calculations were carried out with VASP code  \cite{Kresse1993,Kresse1996,Kresse1996a}, which gives solutions to Density Functional Theory (DFT) using pseudopotentials or the projector-augmented wave (PAW) method.  We used the generalized gradient approximation of Perdew, Burke, and Ernzerhof (PBE-GGA) \cite{Perdew1996} for estimating exchange correlation energies which gives the best agreement to the lattice parameters of $\alpha$ Zr.  Valance configurations of $4s^24p^64d^25s^2$, $2s^22p^4$, and $1s^1$ were considered for Zr, O, and H respectively. All first principles calculations were carried out in $4\times 4\times 4$ Zr supercells (thus 144 Zr atoms), with planes wave cutoff energy of $500\ eV$, and K-point density of $5\times 5\times 5$ which gives very good accuracy in results. Climbing image Nudged Elastic Band (NEB) calculations \cite{Henkelman2000} were carried out to identify the activation energies of these diffusion jumps and the central difference method was used to find the vibration frequencies at energy minima and transition states. Calculated activation energies and vibration frequencies were used to determine the thermally activated hopping rates, while accounting for the Zero Point Energy (ZPE) and quantum tunneling effects using the Semi-Classical Harmonic Transition State Theory (SC-HTST) \cite{Fermann2000,Bhatia2005}. Appendix~\ref{app:SCHTST} shows the procedure in determining the hopping rate using SC-HTST. \citet{Zhang2017} has used this approach to determine the H diffusivity in Zr, which we used to validate the computational method used.

KMC simulations were used to estimate diffusivity of H in bulk Zr with different O concentrations. The smallest repeating cell was used for the KMC models while periodic boundary conditions were incorporated to simulate an infinite crystal. One challenge encountered in KMC calculations was that each pair of nearest neighbour tetrahedral sites acts as ``pseudo-energy basins", due to low activation energy for hopping between them, thus consuming the bulk of the transition steps in the KMC simulations and affecting its efficiency. The mean rate method in \citet{Puchala2010}, provides an analytic solution for this problem by considering pseudo-energy basins as absorbing Markov chains. This helped in finding the modified hopping rates and the mean time of escape from the transient sites which were used in accelerated KMC simulations. Appendix~\ref{app:Markov} shows the method for this modification and validation of the method for one of the pseudo basins encountered in this study. Hopping rates for accelerated KMC were obtained though this method which amalgamates the tetrahedral sites in a pseudo-energy basin to one site.

Diffusivities were calculated for H in pure Zr, and at O concentrations of 0.775\% atm, 1.82\% atm, and 5.88\% atm, for the temperature range $300\ K - 1100\ K$ in steps of $200\ K$. Each of these O concentrations correspond to one O atom in $4\times 4\times 4$, $3\times 3\times 3$, and $2\times 2\times 2$ Zr supercells respectively.  During the study it was assumed that O is distributed uniformly over the Zr lattice, to increase the efficiency of the calculations. At low O concentrations, this assumption will not make a significant difference since O atoms will be spread apart, so their effect on H atoms will not be accumulated. We considered that O only occupies octahedral interstitial sites due to its prohibitively large size for a tetrahedral site \cite{Pemsler1958,Glazoff2013}.  Due to the low solubility of H atoms, they were assumed to diffuse independently from each other. O was considered to be stationary with respect to the H atoms since the diffusivity of O is several orders smaller than that of H. It was also assumed that the Zr lattice does not have any other defects such as vacancies or dislocations to ascertain the effect purely of interstitial O.

\section{Results and Discussion}
We divide this section into two subsections. In the first subsection we discuss results pertaining to diffusivity of H in Pure Zr and using it to validate the methodology followed by comparing with previous work. In the second subsection we discuss the results of how O affects the diffusivity of H. This subsection includes first principle calculations which shows how a single and multiple O atoms affect the behaviour of H occupying interstitial sites in its vicinity and obtain the different hopping rates for these scenarios. Then we used the ascertained hopping rates in KMC simulations to determine H diffusivity for different O concentrations and discuss how O affects the diffusion behavior of H.
\subsection{Hydrogen diffusion in Pure Zr}
\begin{figure}
\includegraphics[width=8.6cm]{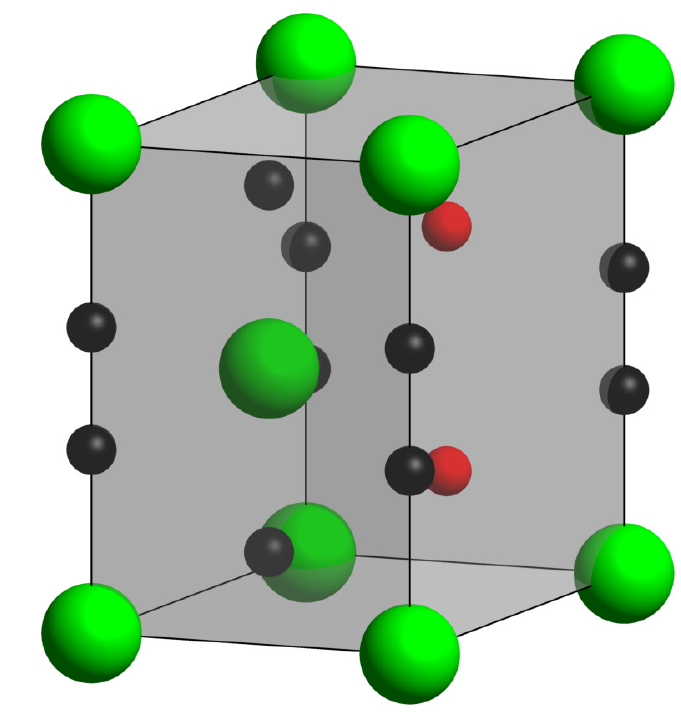}
\caption{\label{fig:unit_cell} Available interstitial site locations for the occupancy of H in a Zr unit cell, where green color  spheres represent the Zr atoms in the hcp lattice, while black and red color spheres represent tetrahedral and octahedral interstitial sites (respectively) available for H occupancy. Some of the tetrahedral sites will be shared between adjacent unit cells.}
\end{figure}

Zr has a conventional unit cell as shown in Fig.~\ref{fig:unit_cell} with four tetrahedral and two octahedral interstitial sites. We found that H favours tetrahedral sites slightly more with an energetic preference of $0.067\ eV$. This value compares well with the values of $0.063\ eV$, $0.057\ eV$, and $0.086\ eV$ reported in \citet{Zhang2017,Domain2002}, and \citet{Burr2012} respectively. The activation energies and the vibration frequencies for the nearest neighbour transitions obtained from NEB calculations and the central difference method,  shown in Table~\ref{tab:act_energy_Zr} and Table~\ref{tab:vib_freq}, also shows good agreement with previous work. With these activation energies and vibration frequencies, hopping rates were determined for each relevant temperature through the procedure given in Appendix~\ref{app:SCHTST}.
\begin{table*}
\caption{\label{tab:act_energy_Zr} Comparison of activation energies of nearest neighbour transitions (in $eV$) in Zr lattice with reported values. `T' denotes tetrahedral sites and `O' denotes octahedral sites.
}
\begin{ruledtabular}
\begin{tabular}{ccccc}
\textrm{Diffusion Step}&
\textrm{Current Work}&
\textrm{\citet{Zhang2017}}&
\textrm{\citet{Domain2002}}&
\textrm{\citet{Christensen2015}}\\
\colrule
T--T &0.132&0.129&0.120&0.129\\
T--O &0.408&0.406&0.410&0.412\\
O--T &0.341&0.346&0.350&0.360\\
O--O &0.403&0.398&0.410&0.427\\
\end{tabular}
\end{ruledtabular}
\end{table*}

\begin{table*}
\caption{\label{tab:vib_freq} Vibration frequencies of the initial and the transition states of each diffusion step. $v_i$ are the vibration frequencies at the initial state, $v_j^{TS}$ are the real-valued vibration frequencies at the transition state, and $v_\pm$ is the imaginary vibration frequency at the transition state. All vibration frequencies are given in $THz$. Values in \citet{Zhang2017} are listed in parentheses for comparison.
}
\begin{ruledtabular}
\begin{tabular}{ccccccc}
Diffusion Step&$v_1$&$v_2$&$v_3$&$v_1^{TS}$&$v_2^{TS}$&$v_\pm$\\
\colrule
T--T &37.08(36.87)&37.08(36.87)&33.62(33.06)&43.79
(43.06)&43.74(43.04)&18.46(17.92)\\
T--O &37.08(36.87)&37.08(36.87)&33.62(33.06)&44.30
(45.80)&41.60(42.66)&16.45(17.59)\\
O--T &23.50(23.32)&21.04(20.84)&21.04(20.84)&44.30
(45.80)&41.60(42.66)&16.45(17.59)\\
O--O &23.50(23.32)&21.04(20.84)&21.04(20.84)&47.39
(47.30)&47.37(47.29)&9.76(9.95)\\
\end{tabular}
\end{ruledtabular}
\end{table*}

\begin{figure}
\includegraphics[width=8.6cm]{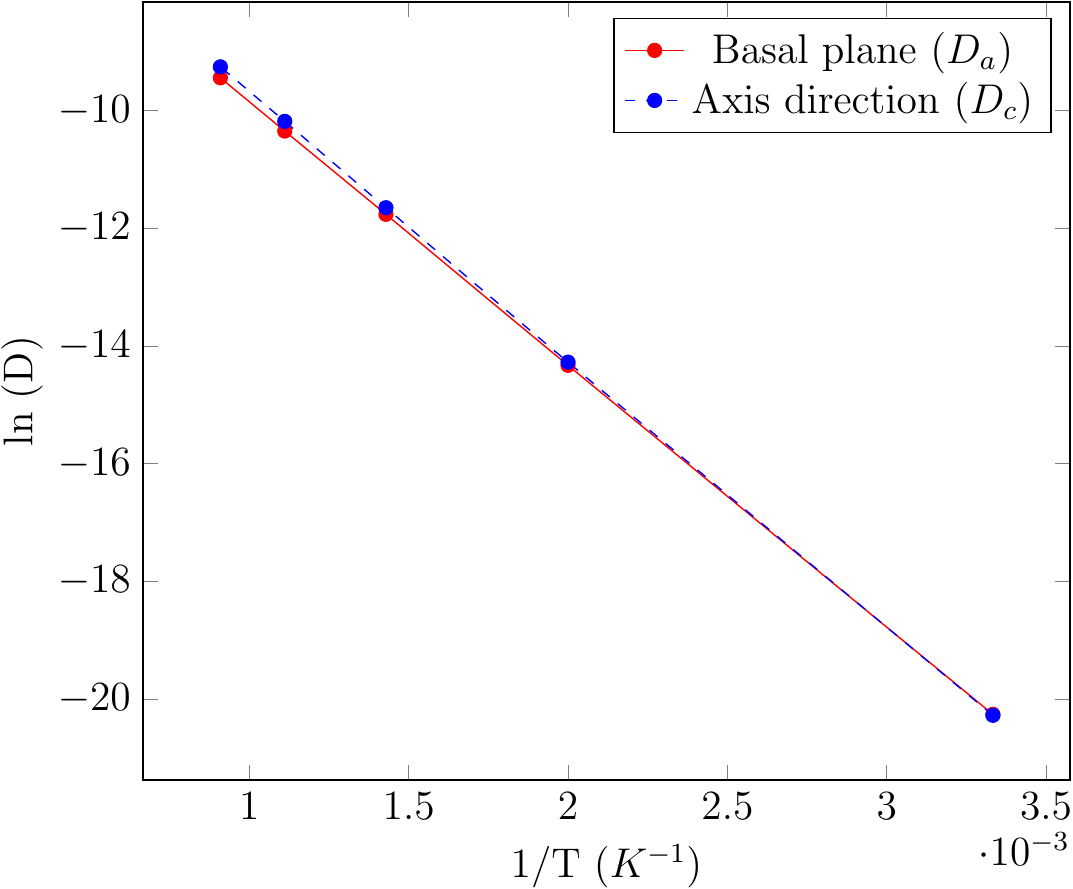}
\caption{\label{fig:diff_Zr} Bulk Diffusivity vs temperature computed from KMC simulations}
\end{figure}

Using these modified rates, KMC simulations were carried out for temperatures ranging between $300\ K-1100\ K$ in steps of $200\ K$, with 100,000 KMC runs for each temperature. Time for each KMC run was selected such that each KMC run lasted more than 1,000,000 diffusion steps in pure Zr. For 100,000 KMC runs, the average diffusivity estimated converged to less than $0.1\%$.
Plotting the logarithm of diffusivity against the inverse of temperature (see Fig.~\ref{fig:diff_Zr}) we obtained the pre-exponent of diffusivity $D_0$ and the activation energy $E_a$ for basal plane and the c-axis direction as shown in Table~\ref{tab:diff_Zr}. These values agree somewhat well with the values available for Zr, where the differences could be attributed to the various defects and impurities in the material as shown in \citet{Zhang2017}. In this study we only looked at the effect of interstitial O on diffusivity, hence we neglected the influence of other impurities and defects.
\begin{table}
\caption{\label{tab:diff_Zr} Diffusivity parameters for H diffusion in Zr
}
\begin{ruledtabular}
\begin{tabular}{ccc}
Direction&$D_0\ (cm^2/w)$&$E_a\ (eV)$\\
\colrule
Basal plane ($D_a$)&$4.535\times 10^{-3}$&0.384\\
C-axis direction ($D_c$)&$5.811\times 10^{-3}$&0.392\\
\end{tabular}
\end{ruledtabular}
\end{table}

From these results we saw that H diffusion in Zr is closely isotropic, especially near the ambient temperature. The anisotropy of H diffusion ($D_c/D_a$, where $D_c$ and $D_a$ are the diffusivities in c-axis and basal plane directions respectively) varies between $0.93-1.18$ in the considered temperature range. Reasons for this variation are discussed in one of our previous papers \citet{Liyanage2018}.

\subsection{Hydrogen diffusion in Zr with interstitial oxygen}
\subsubsection{Effect of oxygen on neighboring interstitial sites}
\begin{figure*}
\subfigure[Naming convention for tetrahedral sites within a basal plane $j$]{\label{fig:index_i_tetra}\includegraphics[width=7cm]{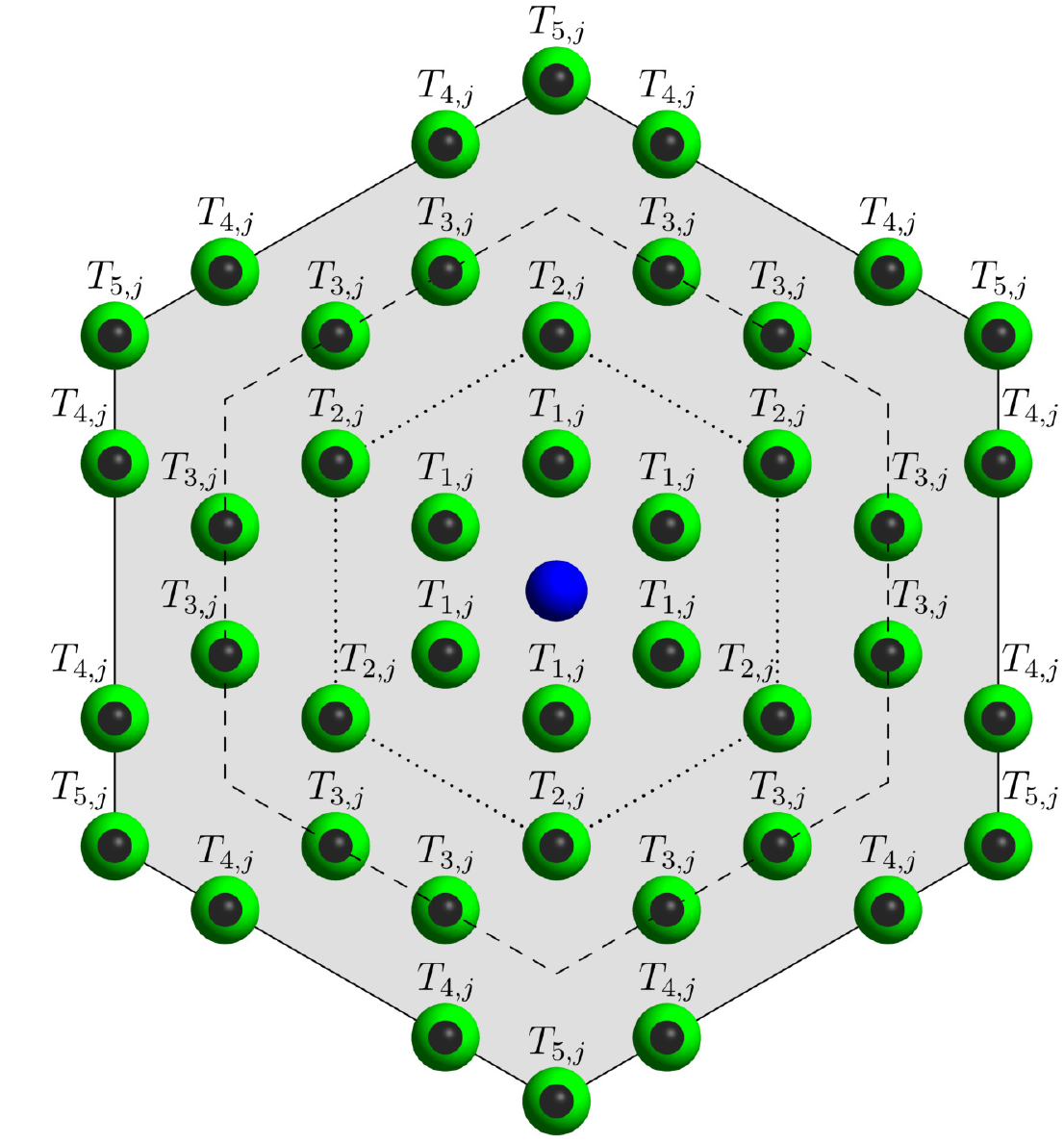}}
\qquad
\qquad
\subfigure[Naming convention for octahedral sites within a basal plane $j$]{\label{fig:index_i_octa}\includegraphics[width=6.6cm]{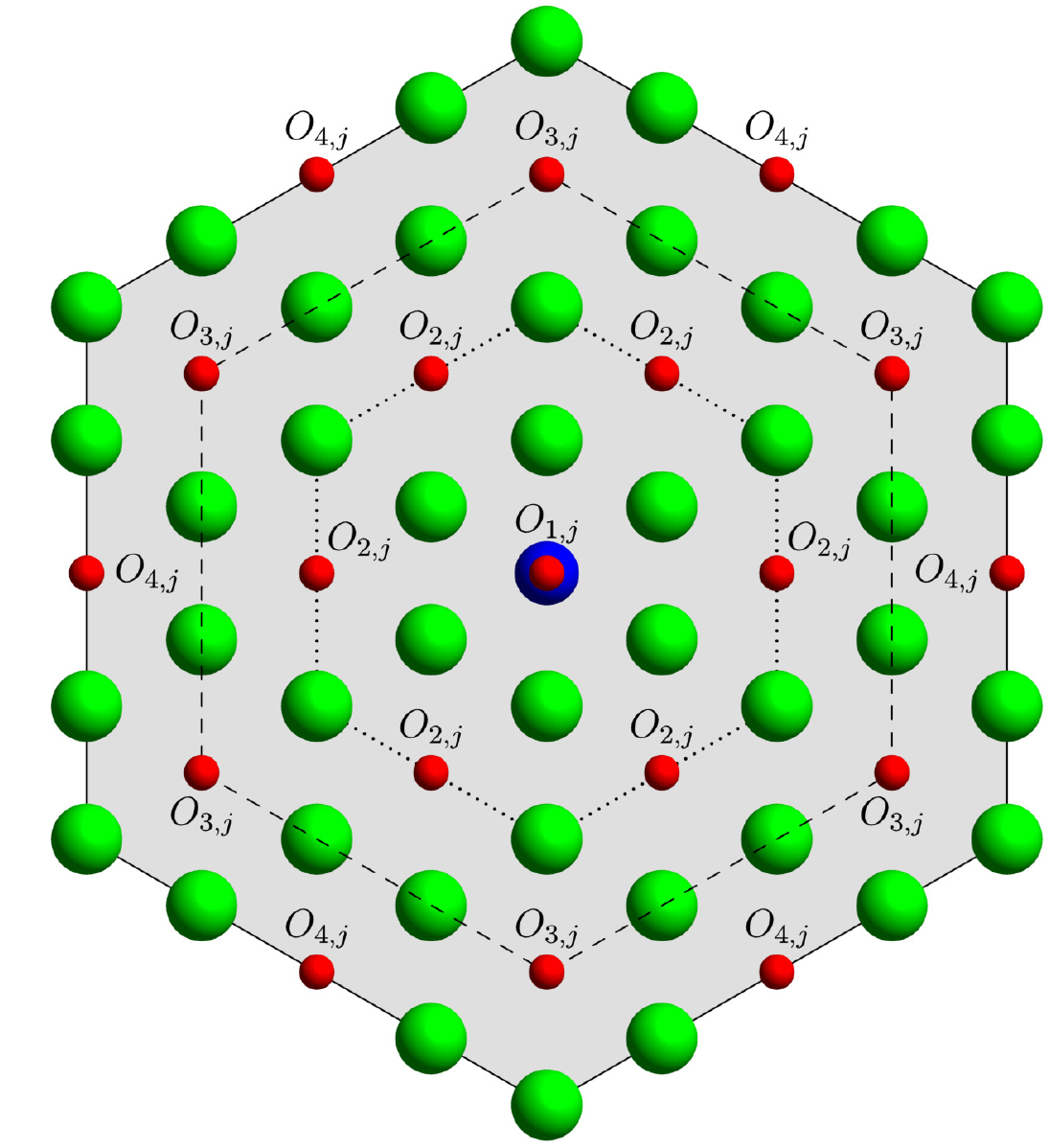}}
\caption{\label{fig:index_i} Variation of index $i$ for tetrahedral and octahedral sites for different O concentrations. Zr, O, tetrahedral sites, and octahedral sites are shown by green, blue, black and red spheres respectively. The region attributed for a single O atom in $4\times 4\times 4$, $3\times 3\times 3$, and $2\times 2\times 2$ Zr supercells are bounded by solid lines, dashed lines, and dotted lines respectively}
\end{figure*}

\begin{figure*}
\subfigure[Naming convention for tetrahedral sites for different basal planes]{\label{fig:index_j_tetra}\includegraphics[width=6cm]{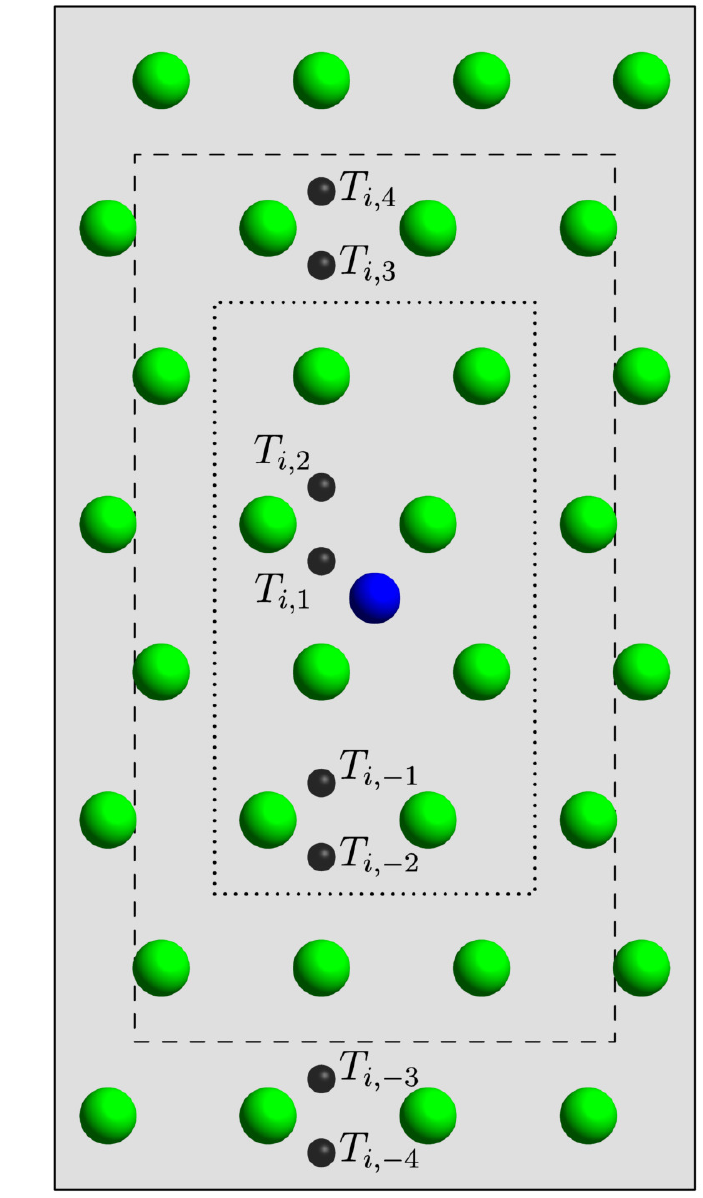}}
\qquad
\qquad
\qquad
\subfigure[Naming convention for octahedral sites for different basal planes]{\label{fig:index_j_octa}\includegraphics[width=6cm]{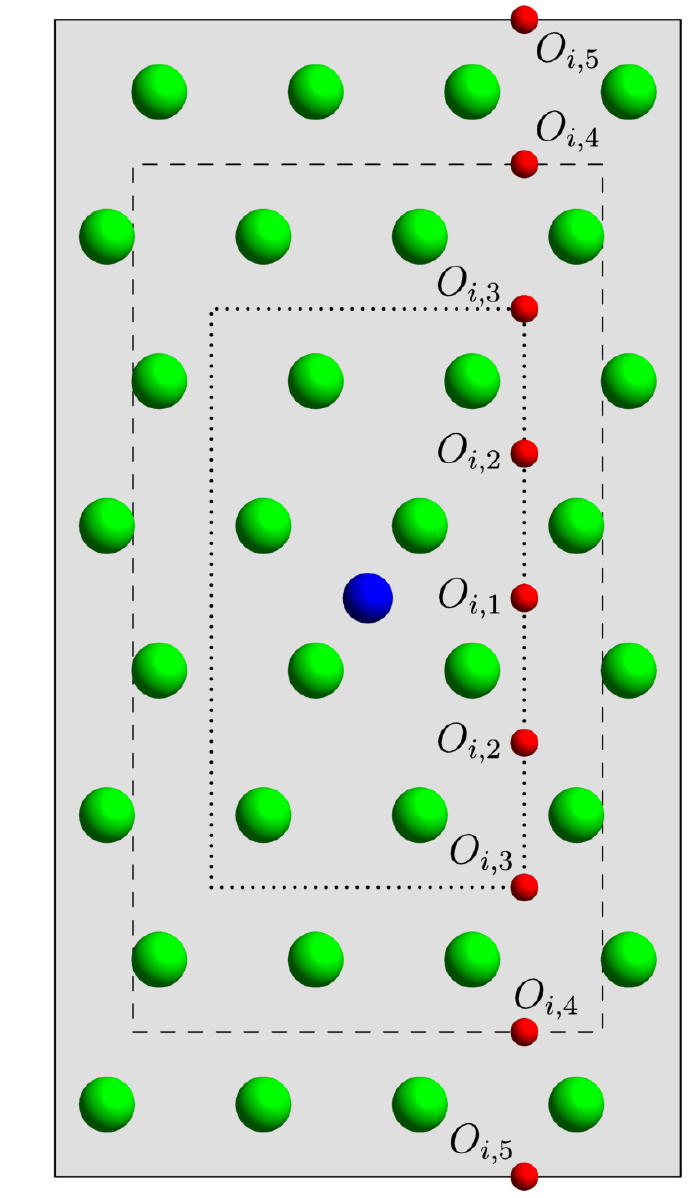}}
\caption{\label{fig:index_j} Variation of index $j$ for tetrahedral and octahedral sites for different O concentrations for any $i$. Colours and bound regions are the same as the previous figure}
\end{figure*}

For the first principle calculations, we used two $4\times 4\times 4$ Zr supercells with one and eight O atoms respectively. In the first system H occupying each interstitial site is only affected by one O atom. For ease of referencing we indexed each distinct interstitial site using the number of steps $i$ to the O in the basal plane and the number of steps $j$ to the plane containing the nearest O along the c-axis. $T_{i,j}$ and $O_{i,j}$ indicate tetrahedral and octahedral interstitial sites respectively. As shown in Fig.~\ref{fig:index_i} and Fig.~\ref{fig:index_j}, $i$ and $j$ indices increases as the distance from O atom increases in basal plane and c-axis direction respectively. Octahedral sites are distributed symmetrically along both sides of O in the c-axis direction while for tetrahedral sites, the variation is not symmetric. So we considered the $j$ indices in direction of nearest neighbour site to be positive and the other direction to be negative.


\begin{figure*}
\subfigure[Tetrahedral sites]{\label{fig:dE_T_single_O}\includegraphics[width=8.6cm]{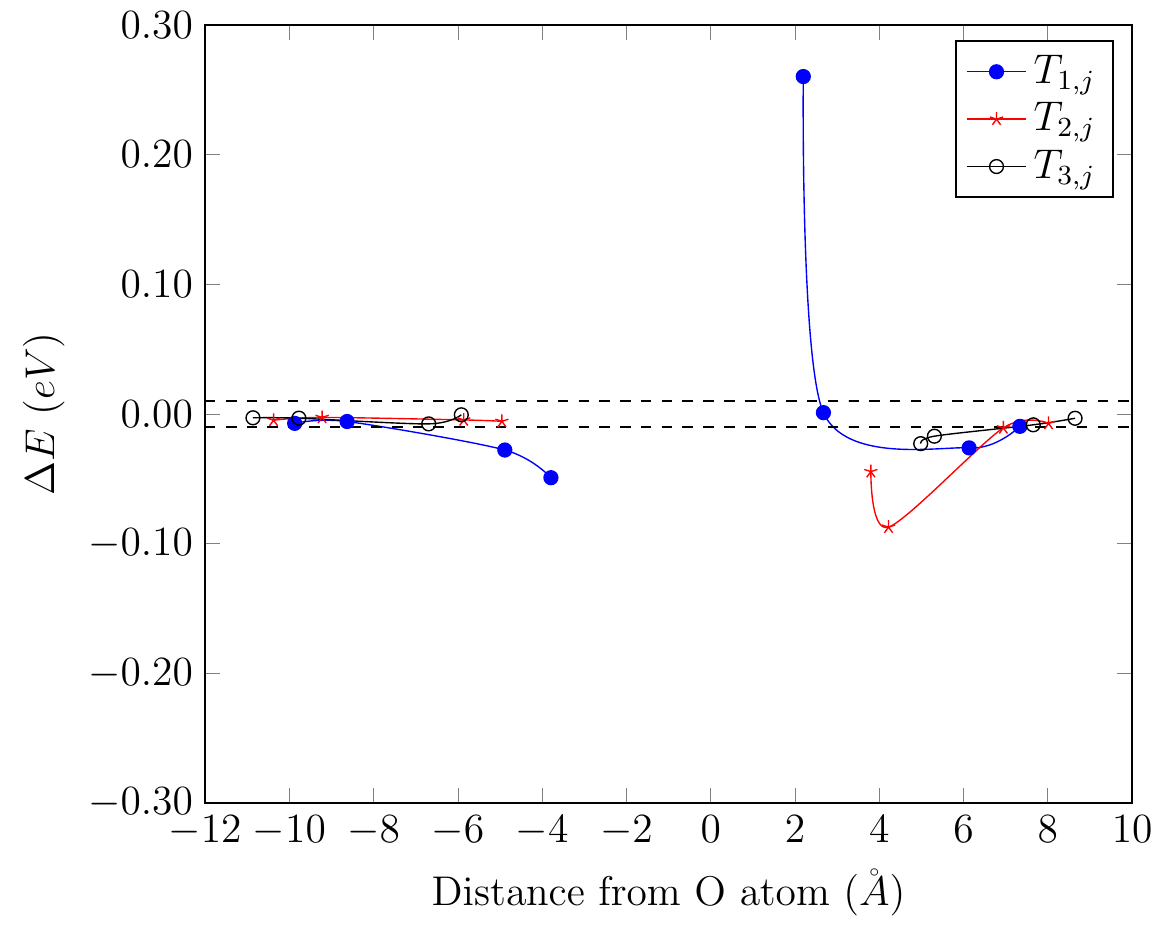}}
\quad
\subfigure[Octahedral sites]{\label{fig:dE_O_single_O}\includegraphics[width=8.6cm]{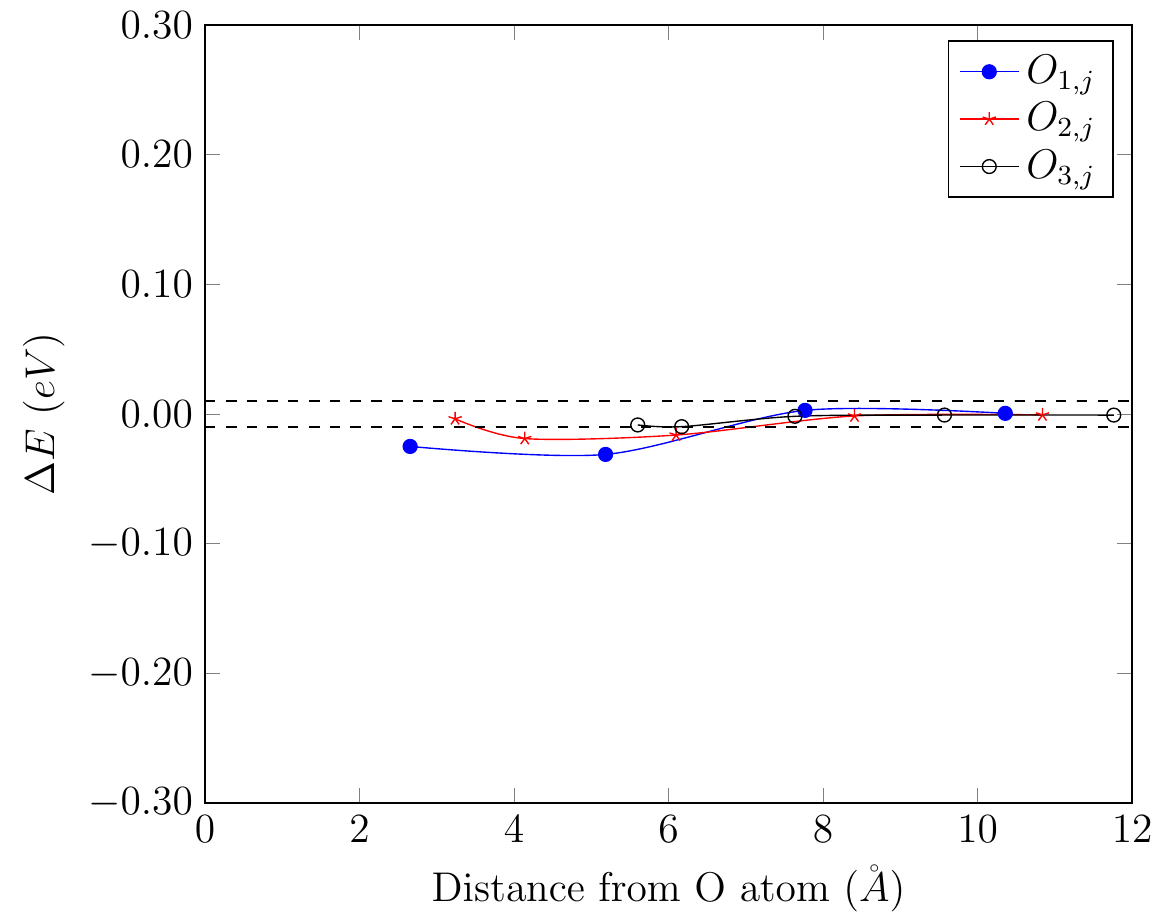}}
\caption{\label{fig:dE_single_O} Variation of $\Delta E$ for interstitial sites when only one O atom is present in the supercell}
\end{figure*}

We calculated the relative change in energy of these interstitial sites when compared to a H occupying the respective interstitial sites in Zr as
\begin{equation}
\Delta E = E_{Zr_{O,H}} - E_{Zr_{O}} - \left( E_{Zr_{H}} - E_{Zr} \right)
\end{equation}
where $E_{Zr_{O,H}}$, $E_{Zr_{O}}$, $E_{Zr_{H}}$, and $E_{Zr}$ are energies of supercell containing both O and H, just O, H in the relevant interstitial site (tetrahedral or octahedral), and no impurities respectively. Fig.~\ref{fig:dE_single_O} shows the variation of $\Delta E$ when only a single O is present. The distances from O atom to tetrahedral sites are taken as positive for $j>0$ and negative for $j<0$. The variation shows a similarity to a classical attractive well, with $\Delta E$ high for very short distances, minimum at some distance, and going to zero at large distances. It shows that the nearest neighboring site to the O atom $T_{1,1}$ has a considerably high energy making it less stable. O is known to block the nearest neighbor tetrahedral sites for H occupancy \cite{Hardie1965} which is consistent with the current results. We also noticed a sharp decrease in energy in the $T_{2,2}$ site making it the most stable location. To simplify the NEB calculations to be preformed, when $\Delta E$ converges to below $0.01\ eV$ we considered the effect of the O atom to be negligible and the behaviour of H at these sites to be similar to that of H in Pure Zr. NEB and the central difference method were used to obtain the activation energies and the vibration frequencies for all the distinct hops. These results were used to find the thermally activated hopping rates from SC-HTST to input in the KMC simulations. One important thing noticed during this step is that the ZPE corrected activation energy for the transition from $T_{1,1}$ to $T_{1,2}$ becomes negative, which means that $T_{1,1}$ is not a stable occupancy location and will automatically relax to the $T_{1,2}$ site, making $T_{1,1}$ a part of the $T_{1,2}$ energy basin.

\begin{figure}
\includegraphics[width=5cm]{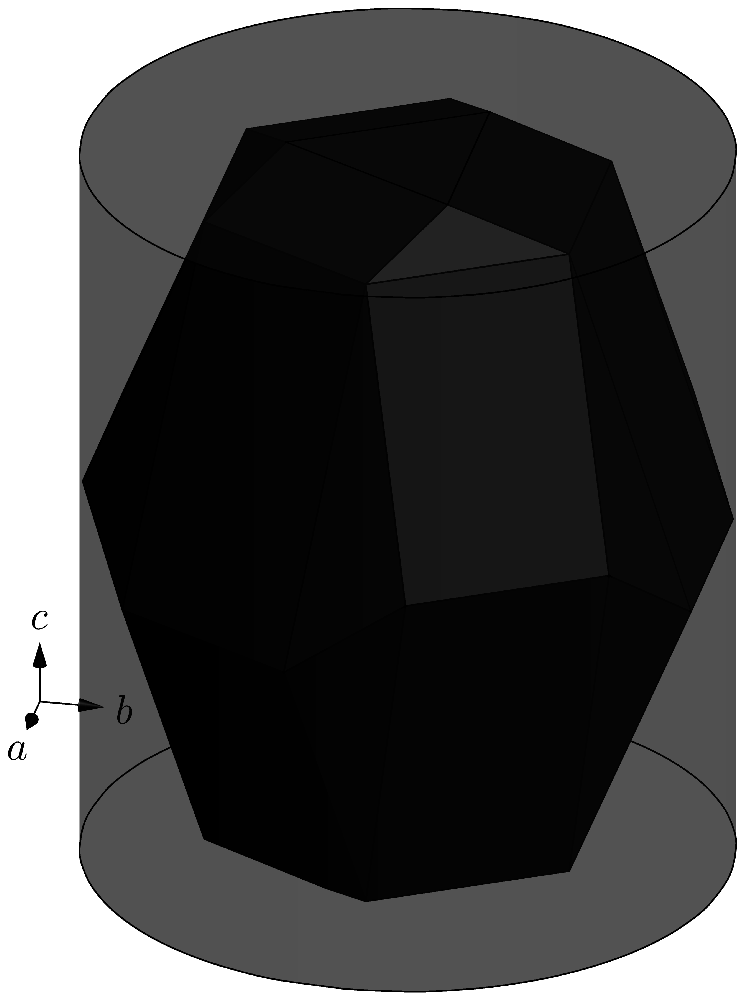}
\caption{\label{fig:cylinder} Region affected by a single O atom which can be simplified to the cylindrical region for ease of representation}
\end{figure}

\begin{figure*}
\subfigure[$4\times 4\times 4$ supercell]{\label{fig:regions_4}\includegraphics[width=6cm]{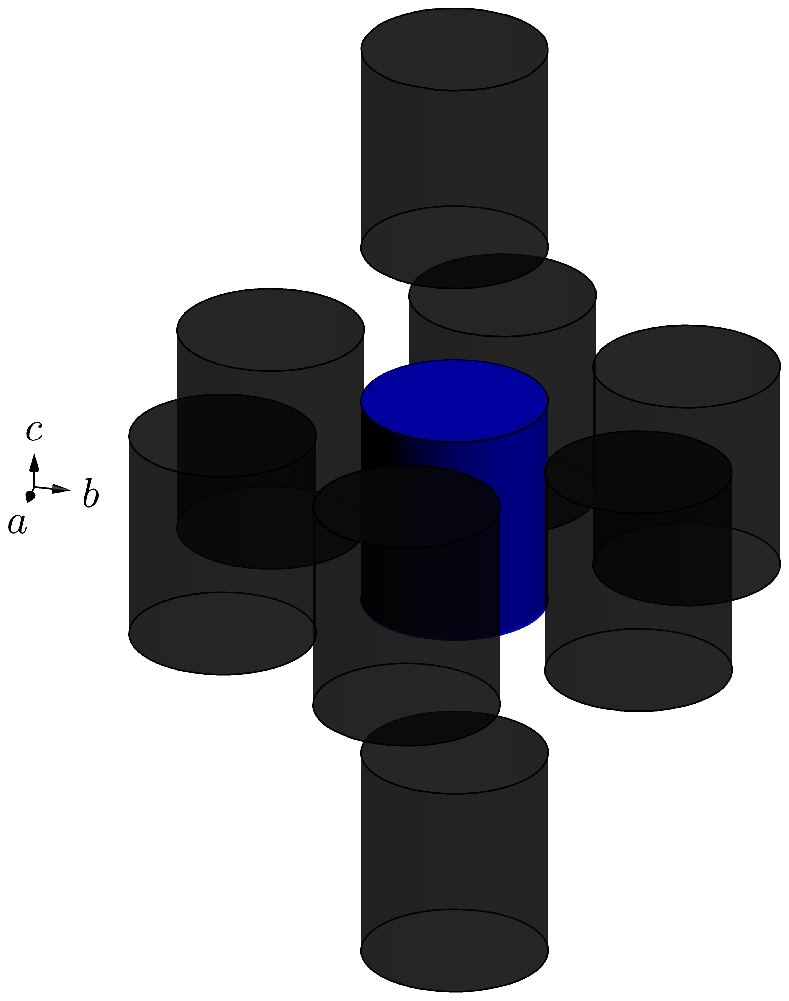}}
\qquad
\qquad
\qquad
\subfigure[$3\times 3\times 3$ supercell]{\label{fig:regions_3}\includegraphics[width=6cm]{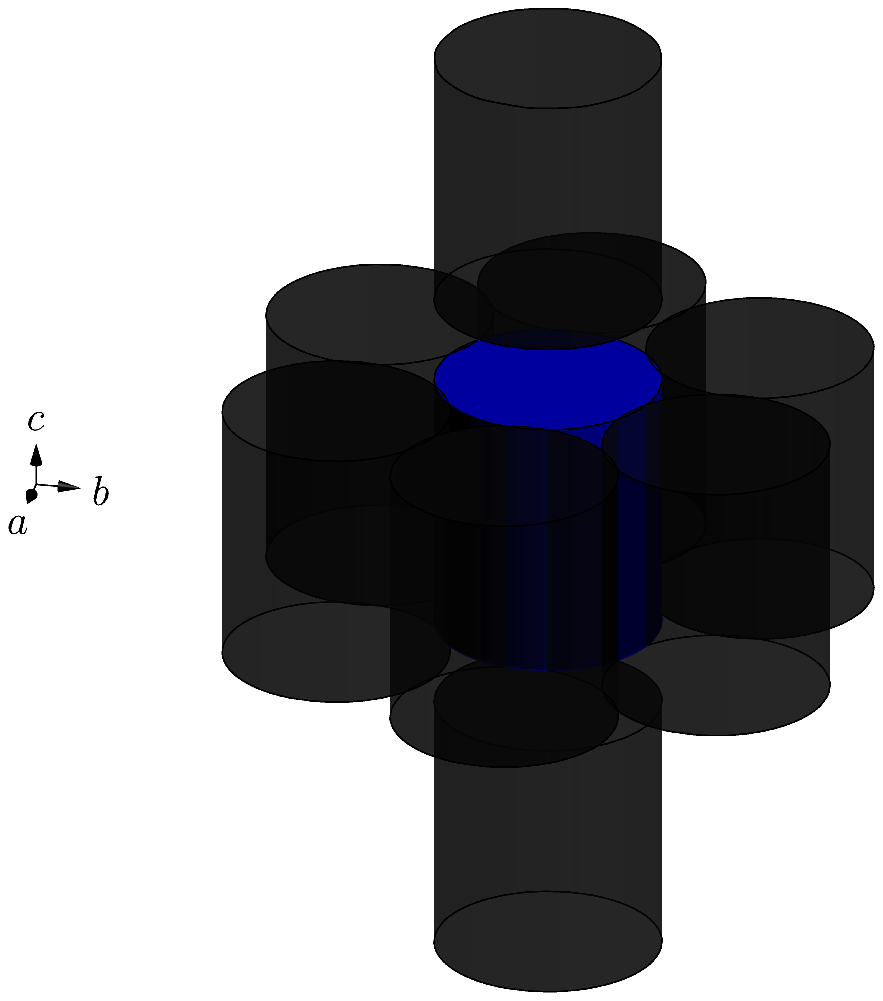}}\
\caption{\label{fig:regions} Arrangement of neighbouring regions by O atoms}
\end{figure*}

\begin{figure*}
\subfigure[Tetrahedral sites]{\label{fig:dE_T_multiple_O}\includegraphics[width=8.6cm]{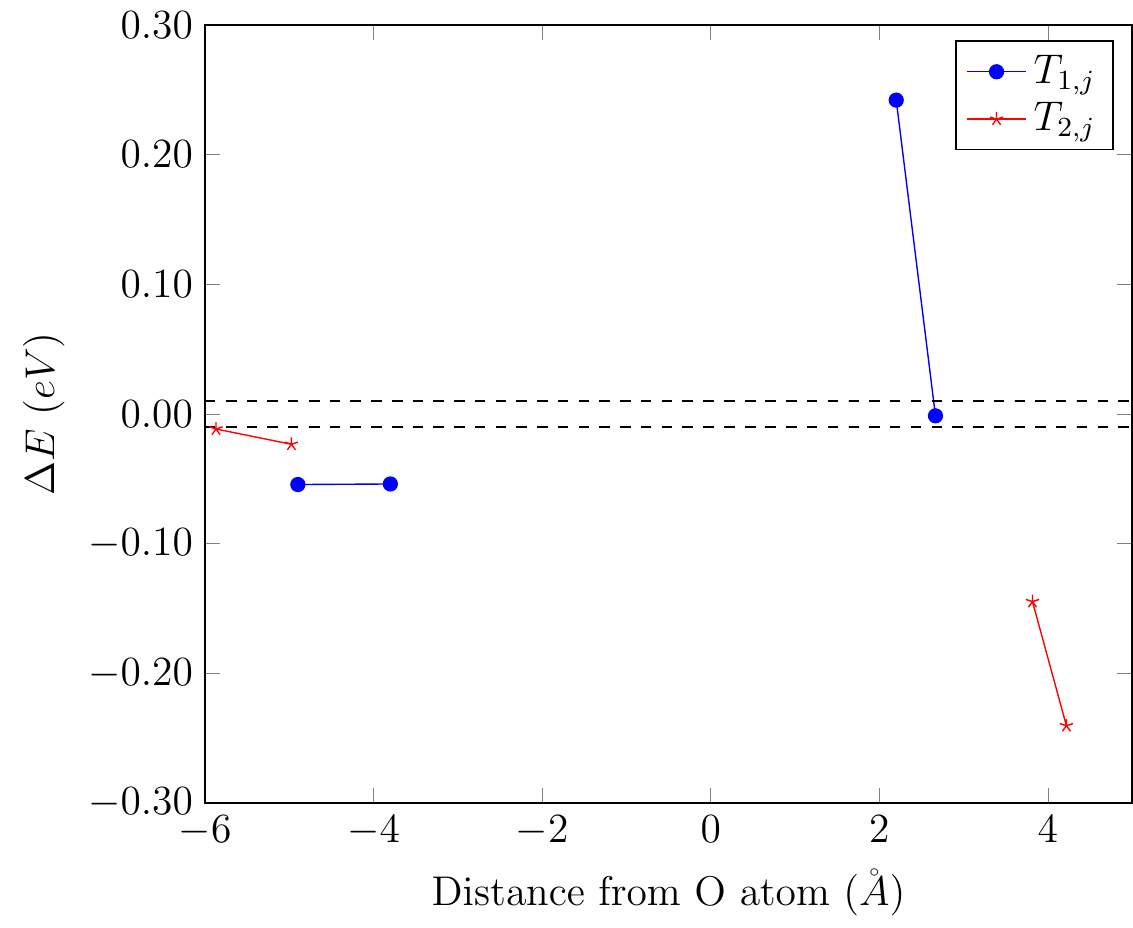}}
\subfigure[Octahedral sites]{\label{fig:dE_O_multiple_O}\includegraphics[width=8.6cm]{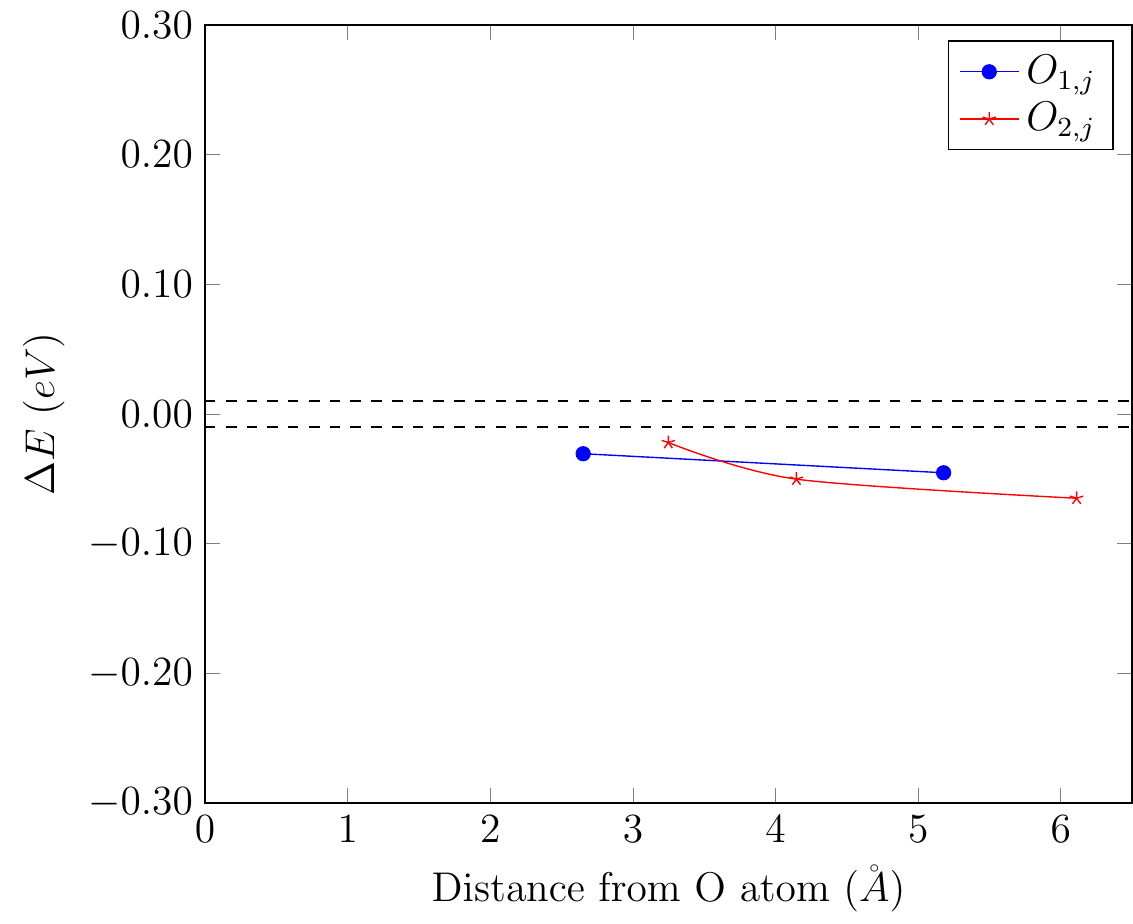}}
\caption{\label{fig:dE_multiple_O} Variation of $\Delta E$ for interstitial sites when eight O atoms are present in the supercell}
\end{figure*}

The region affected by an O atom is as shown in Fig.~\ref{fig:cylinder}, where the polyhedron encompasses all the site of which $\Delta E>0.1\ eV$. This can be simplified to a cylindrical region for ease of representation. If we look at the distribution of affected regions when an O atom is present in each of $4\times 4\times 4$ and $3\times 3\times 3$ Zr unit cells, the arrangement of neighbouring regions are shown in Fig.~\ref{fig:regions}. For the $4\times 4\times 4$ system, the affected regions are spread apart where each interstitial site is at most affected by a single O atom, allowing us to use the calculated hopping rates to simulate H diffusivity for this O concentration. In the $3\times 3\times 3$ system we can see that the that the regions are touching each other in the basal plane direction. But this will only affect $T_{3,1}$ interstitial sites in the perimeter of the region, and hence we still can use the hopping rates calculated without a significant effect.
For the second system we considered eight O atoms in a $4\times 4\times 4$ Zr supercell (equivalent to one O atom in $2\times 2\times 2$ supercell), where from previous results we can see that some interstitial sites will be affected by multiple O atoms. Similar to the above system we looked at the variation of $\Delta E$ for the distinct interstitial locations. As Fig.~\ref{fig:dE_multiple_O} shows, we can see that all the interstitial sites are affected by the presence of O atoms. Similar to the single O system, we can see that $T_{1,1}$ sites are quite unstable. $T_{2,2}$ sites are still the most stable sites, with the energetic favorability having increased further. Activation energies and vibration frequencies obtained from NEB and central difference method for each distinct transition for both systems are provided in the supplemental material which were used to determine the thermally activated hopping rates for each transition.
\subsubsection{Effect of oxygen on diffusivity}
\begin{figure*}
\subfigure[Basal plane]{\label{fig:diff_O_basal}\includegraphics[width=8.6cm]{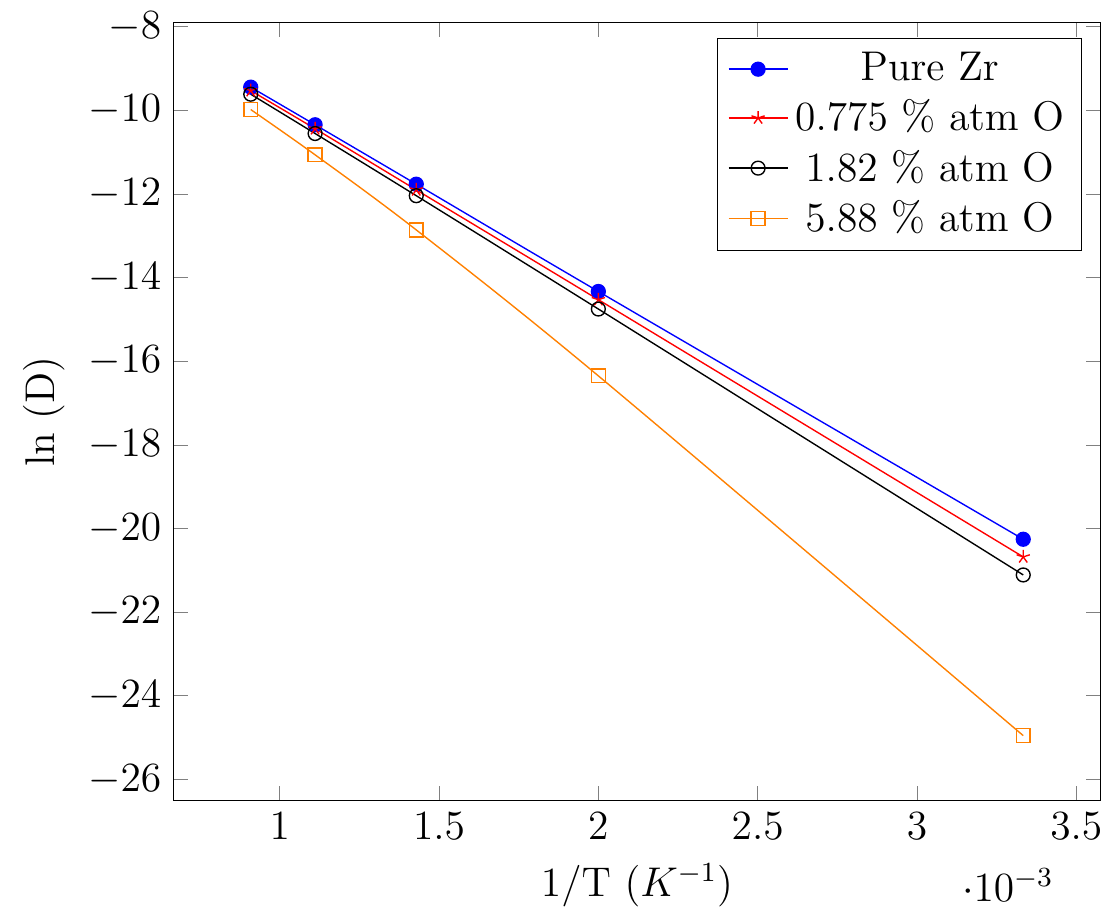}}
\subfigure[C-axis direction]{\label{fig:diff_O_axis}\includegraphics[width=8.6cm]{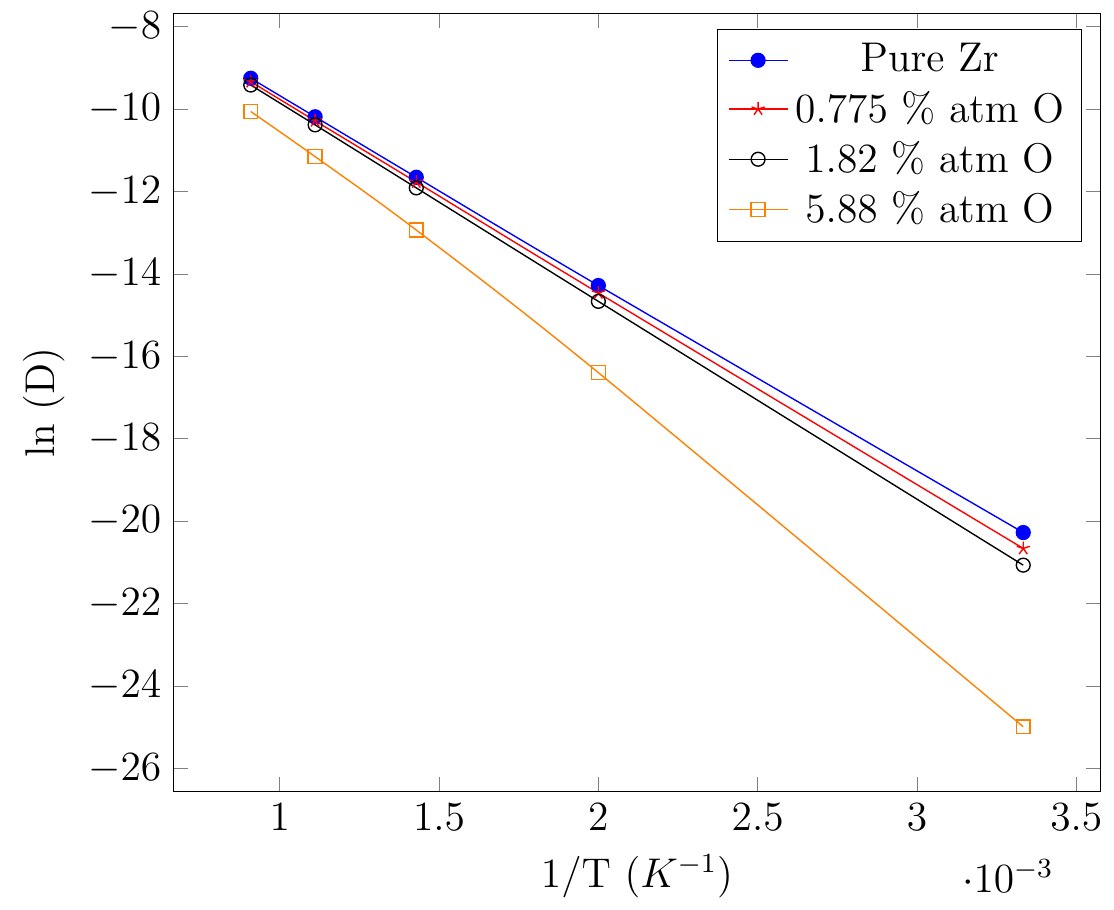}}
\caption{\label{fig:diff_O} Variation of diffusivity for with O concentration}
\end{figure*}

Using the hopping rates determined, we performed KMC simulations for the three difference O concentrations. Variations of diffusivity for each concentrations are as shown in Fig.~\ref{fig:diff_O} for basal plane and c-axis directions. As can be seen from this graph when the O concentration increases the H diffusivity reduces, proving the hypothesis of the current study. We can see that at the highest O concentration the relationships are not linear deviating from the Arrhenius form
\begin{equation}
    D = D_0 \exp \left(-E_a/k_BT\right),
\end{equation}
while for low concentrations they become close to the Arrhenius form.

\begin{figure*}
\subfigure[Basal plane]{\label{fig:change_basal}\includegraphics[width=8.6cm]{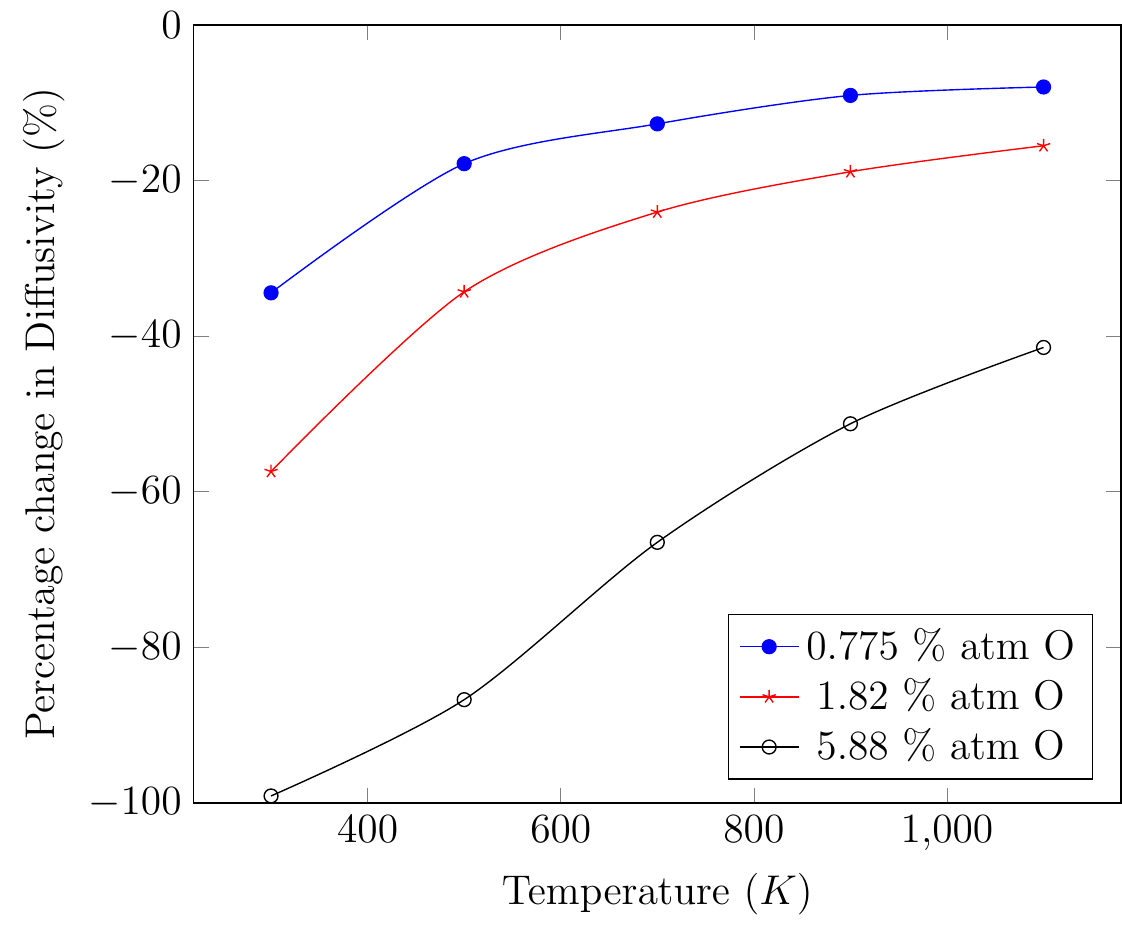}}
\subfigure[C-axis direction]{\label{fig:change_axis}\includegraphics[width=8.6cm]{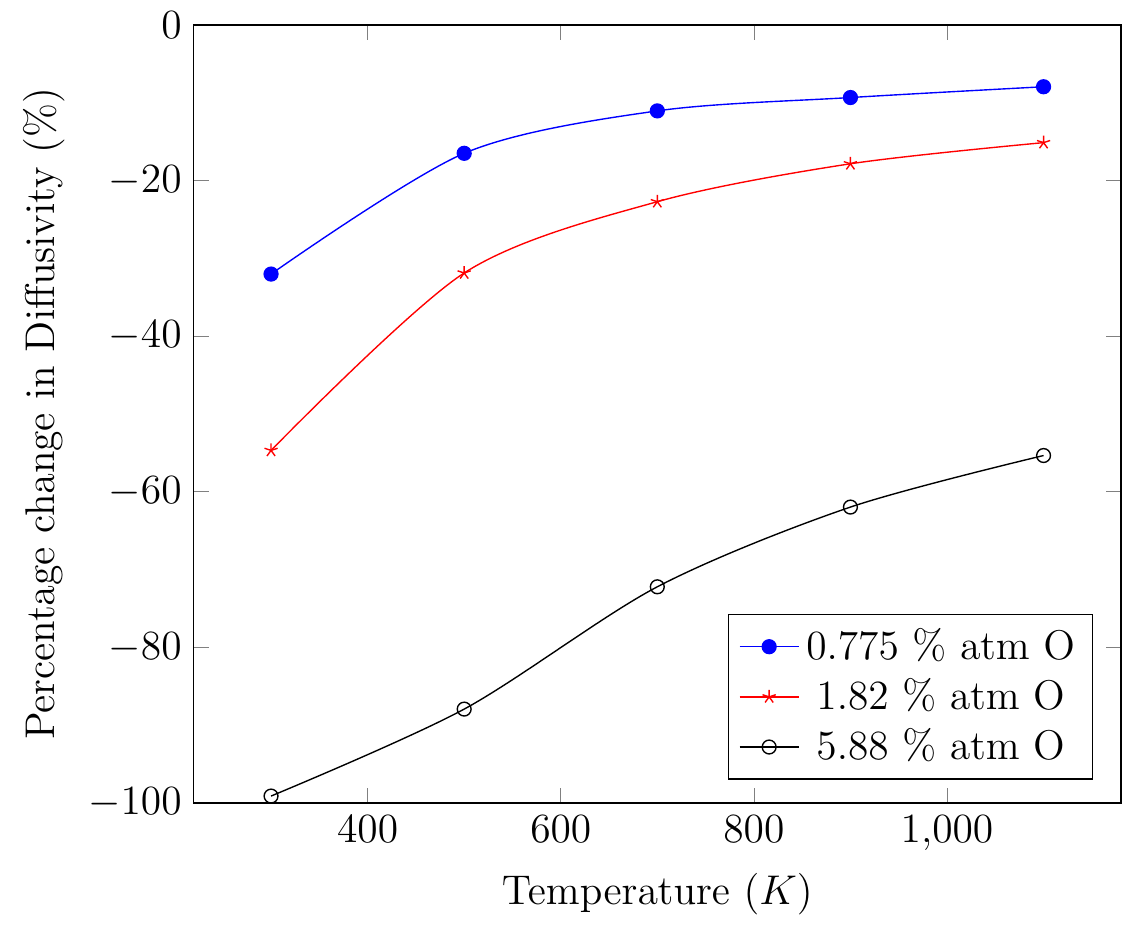}}
\caption{\label{fig:change} Percentage change in diffusivity compared to that of Pure Zr}
\end{figure*}

We calculated the diffusivity parameters $D_0$ and $E_a$ for H diffusion as shown in Table~\ref{tab:diff_O_basal} and Table~\ref{tab:diff_O_axis} for basal plane and c-axis directions respectively. If we look at the change in diffusivity with respect to the diffusivity in pure Zr (see Fig.~\ref{fig:change}), we can see that the decrease in diffusivity is influenced by both the temperature and the O concentration. This is evident by the higher activation energy through increased O concentration.
\begin{table}
\caption{\label{tab:diff_O_basal} Diffusivity parameters for H diffusion in the basal plane direction for different O concentrations
}
\begin{ruledtabular}
\begin{tabular}{ccc}
O concentration&$D_0\ (cm^2/w)$&$E_a\ (eV)$\\
\colrule
Pure Zr&$4.535\times 10^{-3}$&0.384\\
0.775\% atm O&$4.828\times 10^{-3}$&0.397\\
1.82\% atm O&$5.089\times 10^{-3}$&0.409\\
5.88\% atm O&$15.945\times 10^{-3}$&0.535\\
\end{tabular}
\end{ruledtabular}
\end{table}

\begin{table}
\caption{\label{tab:diff_O_axis} Diffusivity parameters for H diffusion in the c-axis direction for different O concentrations
}
\begin{ruledtabular}
\begin{tabular}{ccc}
O concentration&$D_0\ (cm^2/w)$&$E_a\ (eV)$\\
\colrule
Pure Zr&$5.811\times 10^{-3}$&0.392\\
0.775\% atm O&$6.109\times 10^{-3}$&0.403\\
1.82\% atm O&$6.427\times 10^{-3}$&0.414\\
5.88\% atm O&$14.402\times 10^{-3}$&0.533\\
\end{tabular}
\end{ruledtabular}
\end{table}

To identify the reasons for this variation we looked at the diffusion paths taken by H atoms for each of the O concentrations at $300\ K$ where the highest difference to the diffusivity is observed. At the lowest O concentration (0.775\% atm), about 60\% of the diffusion steps occur between interstitial sites unaffected by the O atom. This can be expected since the majority of the interstitial sites at this concentration are out of the influence from O atoms (more than 70\% of both tetrahedral and octahedral sites respectively). At 1.82\% atm O concentration, only 20\% the transitions are among these sites, even though still about 45\% of tetrahedral sites and 35\% of the octahedral sites are not under the influence of O atoms. This shows that H prefers the influenced region for occupancy making these regions trapping sites for H atoms.

For the given time interval, the number of diffusion jumps reduces from that of pure Zr by approximately 26\%, 42\%, and 94\% as the O concentration increases. This compares well to the reduction in diffusivities of approximately 35\%, 57\%, and 99\% respectively. This is as expected since it takes more energy to remove an atom from a more stable location. The additional reduction is due to the flickering (back and forth movement between two or more neighboring sites) events happening between the sites in the affected region which is very much noticeable in the system with the highest O concentration.

\begin{figure}
\includegraphics[width=8.6cm]{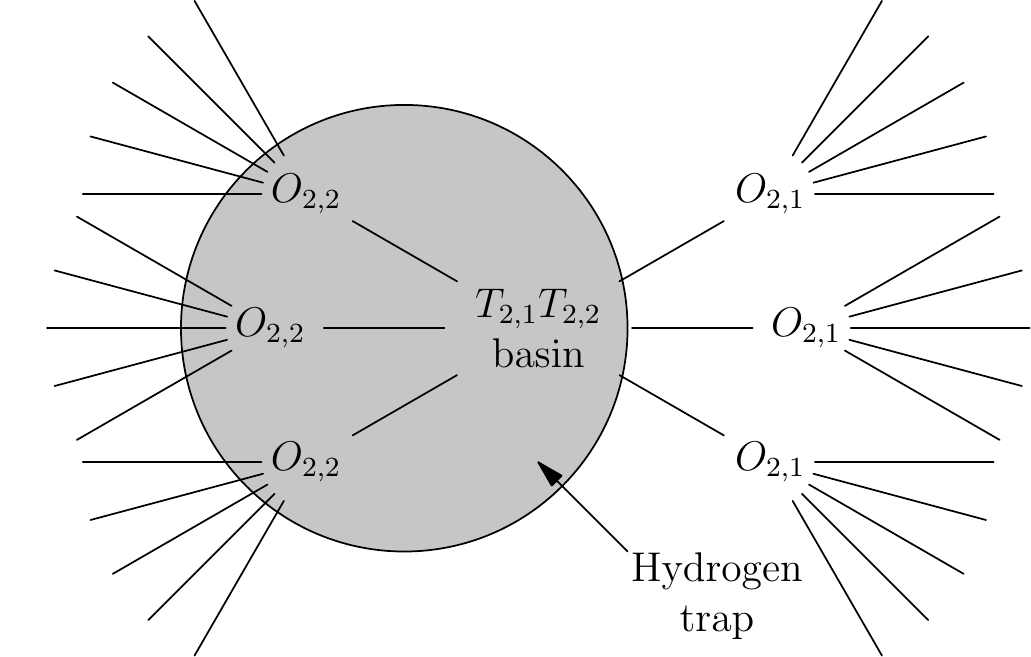}
\caption{\label{fig:basin} Trap formed by pseudo basin $T_{2,1}T_{2,2}$ and $O_{2,2}$ sites. H atoms arriving in these traps will flicker between the sites before escaping, causing a reduction in diffusivity}
\end{figure}

In this system, transitions are mainly among $O_{2,2}$ sites and pseudo basins formed by adjacent $T_{2,1}$ and $T_{2,2}$ sites, accounting for more than 80\% of the transitions. Each set of this pseudo basin and the three nearest neighbour $O_{2,2}$ sites forms a sink for H atoms where they are trapped between these sites for several diffusion steps before escaping it (see Fig.~\ref{fig:basin}).  This effect is similar to the pseudo basins formed by nearest neighbour tetrahedral sites, but to a smaller extent. This causes the diffusivity at this concentration to reduce further.

\section{Conclusions and Future Direction}
We looked at diffusivity of H in Zr with and without the influence of O using a combination of first principles methods and KMC simulation. The first principles calculations have been successfully used to determine the hopping rates for individual diffusion steps, validated from previous work carried out for H diffusion in pure Zr. This study proves the hypothesis that H diffusion in Zr  is decreased by the presence of interstitial O and further decrease as the O concentration increases. We looked at the paths taken by H atoms and found that in addition to the decrease in hopping rates, H traps are formed by amalgamation of occupancy sites, decreasing the diffusivity.

While this study proves that diffusivity of H in Zr is significantly affected by O, it was carried out under ideal conditions. For a more general study the change in diffusivity can be determined while incorporating other impurities and defects. The diffusivity values obtained from this study can be used in a precipitation model to look at the effect it will have on the size of the hydrides formed with varying O concentrations.

Here we have only considered the effects of low to moderate O concentrations whereas O has very high solubility levels which can go up to 30\% atm. This study can also be extended to high O concentrations, but at these concentrations the uniform distribution of O interstitials may not be a valid assumption. A non-uniform distribution will increase the computational effort considerably, as a variety of different O arrangements need to considered.
\begin{acknowledgments}
The authors gratefully acknowledge funding from NSERC's Discovery Grants program.
\end{acknowledgments}
\appendix
\section{\label{app:SCHTST}Semi-Classical Harmonic Transition State Theory}

SC-HTST formulated by \citet{Fermann2000} only requires the activation energy and vibration frequencies of initial and transition state. Climbing image NEB was used to find the activation energy and the central difference method was used to determine the vibration frequencies by decoupling the H atoms with the remaining atoms. The hopping rate from SC-HTST ($k^{SC-HTST}$) can be calculated as
\begin{equation}
k^{SC-HTST} = k^{HTST} \Gamma (T)
\end{equation}
where $k^{HTST}$ is the harmonic transition state theory hopping rate and $\Gamma (T)$ is the quantum tunneling correction. $k^{HTST}$ can be calculated as
\begin{equation}
k^{HTST} = \frac{\prod_{i = 1}^{3} v_{i}f(hv_{i}/2k_BT)}{\prod_{j = 1}^{2} v_{j}^{TS}f(hv_{j}^{TS}/2k_BT)} \exp{(-E_a/k_BT)}
\end{equation}
where, $h$ is the Planck's constant, $v_{i}$ and $v_{j}^{TS}$ are the vibration frequencies associated with the initial state and the transition state respectively, $k_B$ is the Boltzmann constant, $T$ is the temperature, and the function $f(x)=sinh(x)/x$. This already includes the effect of zero-point energy (ZPE) on the classical activation energy $E_a$. The quantum tunneling is a temperature dependant multiplier to be determined as
\begin{equation}
\Gamma (T) = \frac{\exp{(E_{ZP}/k_BT)}}{1+\exp{(2\theta_0)}}+\frac{1}{2}\int_{-\infty}^{\theta_0} \sech^2{\theta} \exp{\left( \frac{hv_{\pm}\theta}{\pi k_BT}\right)} d\theta
\end{equation}
where $\theta_0=(\pi E_{ZP})/(hv_{\pm})$ and $v_{\pm}$ is the imaginary vibration frequency at the transition state. $E_{ZP}$ is the ZPE corrected activation energy to be calculated as
\begin{equation}
E_{ZP} = E_a - \frac{1}{2}\sum_{i = 1}^{3} hv_{i} + \frac{1}{2}\sum_{j = 1}^{2} hv_{j}^{TS}
\end{equation}
\section{\label{app:Markov}Absorbing Markov Chains}
As discussed one major issue faced during KMC simulation of Zr and other hcp metals, is the back and forth movement between nearest neighbor tetrahedral sites. While there are several methods available to solve this problem, we followed the mean rate method shown in \citet{Puchala2010} where the tetrahedral sites are considered as transient states, and the octahedral sites are considered as absorbing states in a Markov chain for each pseudo-energy basin. Using this method, the rate of escape from the pseudo-energy basin to its nearest neighbor sites are determined.

Initially the transition probability matrix $\Bar{\Bar T}$ is created with entries
\begin{equation}
T_{ij}=\frac{R_{i\rightarrow j}}{\sum_k R_{i\rightarrow k}}
\end{equation}
where $R_{i\rightarrow j}$ is the hopping rate from transient state $i$ to $j$ and $k$ are all the transient and absorbing states available to transition from state $i$. This can be also written as
\begin{equation}
T_{ij}=\tau_i^1 R_{i\rightarrow j}
\end{equation}
where $\tau_i^1=\sum_k R_{i\rightarrow k}$ is the mean residence time at state $i$. The probability of still being in each transient state after $m$ jumps, the occupational probability vector can be determined as
\begin{equation}
\Bar \Theta (m) = \Bar{\Bar T}^m \Bar \Theta(0)
\end{equation}
where $\Bar \Theta(0)$ gives the occupancy at the start with entries,
\begin{equation}
\\\Theta_i(0) = \left\{
  \begin{array}{ll}
    1 & \text{if }i=s_{init}\\
    0 & \text{otherwise}
  \end{array}
\right.
\end{equation}
with $s_init$ being the initial occupancy site.
By summing the occupational probability over all $m$ we get the total occupational probabilty vector as
\begin{equation}
\Bar \Theta^{tot} = \sum_{m=0}^\infty \Bar{\Bar T}^m \Bar \Theta(0) = \left(1-\Bar{\Bar T}\right)^{-1}\Theta(0)
\end{equation}
From this, the mean residence time at each transient state can be determined as
\begin{equation}
\tau_i =\tau_i^1 \Theta_i^{tot}
\end{equation}

The rates for escaping into each of the absorbing state can be modified via
\begin{equation}
\langle R_{i\rightarrow j} \rangle =\frac{\tau_i}{\sum_k \tau_k} R_{i\rightarrow j}
\end{equation}
with $k$ summed over all transient states.

For nearest neighbor tetrahedral sites pseudo-energy basins the initial occupancy site can be either of the two sites. In the solutions it didn't give an appreciable difference to the rates based on the initial occupancy location. Hence, we calculated the modified rates for both cases and then took the average rate to determine the rates to be used in KMC simulations.

\begin{figure}
\includegraphics[width=8.6cm]{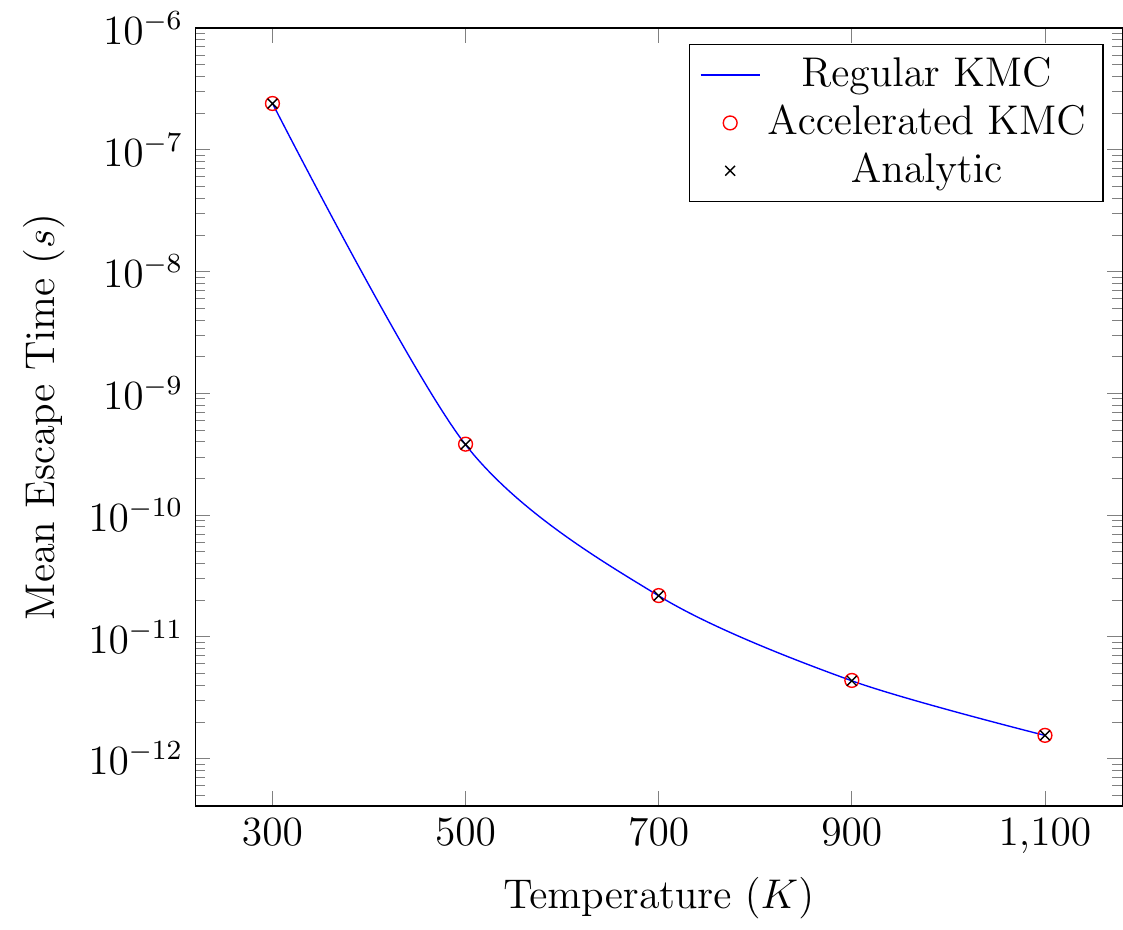}
\caption{\label{fig:escape_time} Mean escape time from the pseudo-energy basin formed by nearest neighbour $T_{2,1}$ and $T_{2,2}$ sites}
\end{figure}

\begin{figure}
\includegraphics[width=8.6cm]{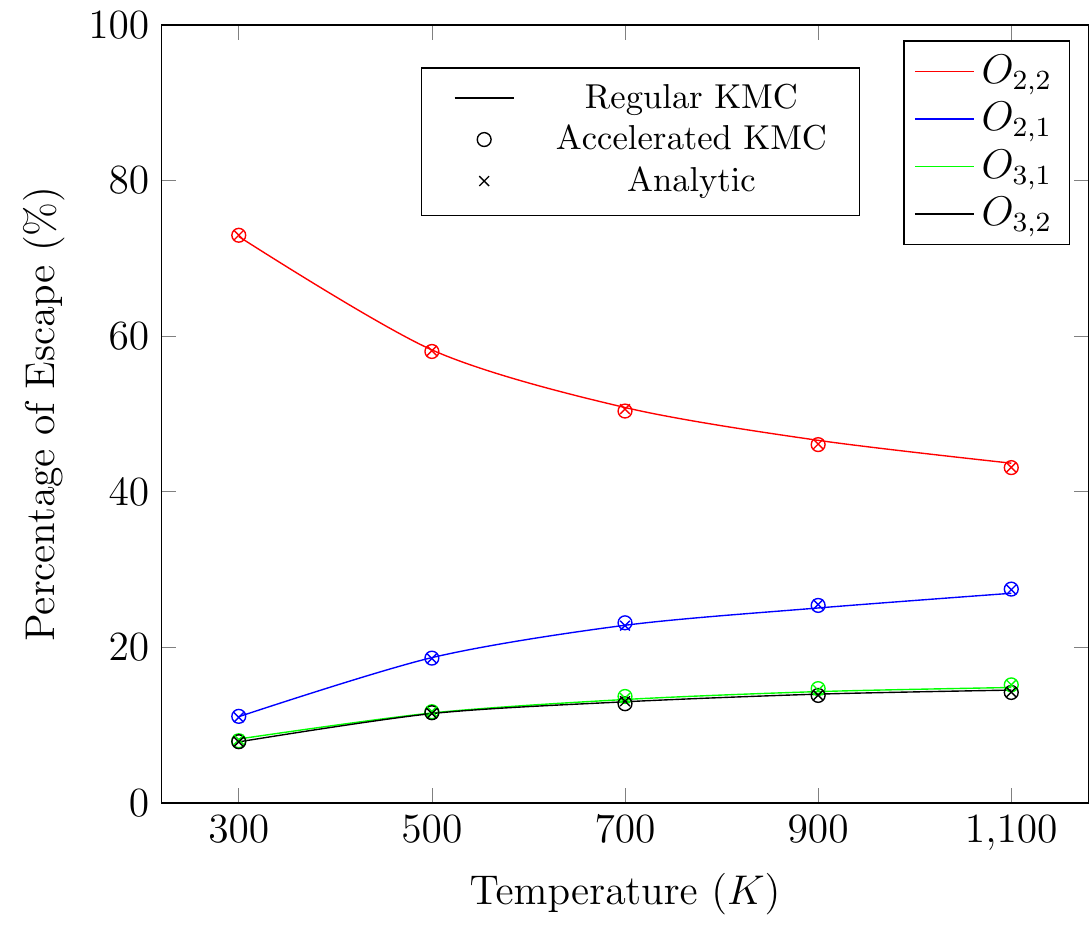}
\caption{\label{fig:escape_prob} Probability of exiting along possible paths from the pseudo-energy basin formed by nearest neighbour $T_{2,1}$ and $T_{2,2}$ sites. Solid lines shows the results from regular KMC, the $\circ$ shows the results from accelerated KMC, and the $\times$ shows the analytic solution. Note that the two lowest lines fall virtually atop each other}
\end{figure}

To check the validity of this method, we looked at the variation of average basin exit time and the probability of taking each of the escape path for the regular KMC, accelerated KMC, and the analytic solution by looking at a pseudo basin formed by $T_{2,1}$ and $T_{2,2}$ sites for 100,000 KMC runs. A H atom in $T_{2,1}$ has the possibility to exit to one $O_{3,1}$ site and two $O_{2,1}$ sites, while $T_{2,2}$ has the possibility to exit to one $O_{3,2}$ site and two $O_{2,2}$ sites. As seen by Fig.~\ref{fig:escape_time} and \ref{fig:escape_prob} we can see that the three methods gives really similar results validating this method.

\nocite{supp1}
\bibliography{Oxygen_on_diffusion_of_Hydrogen}% Produces the bibliography via BibTeX.

\end{document}